\documentclass[aps,pra,twocolumn,showpacs]{revtex4-1}
\usepackage{amsmath,amssymb,mathrsfs}
\usepackage{graphicx}
\usepackage{amsmath}
\usepackage{epsfig}
\usepackage[usenames]{color}
\usepackage{amsmath}
\usepackage{amssymb}
\usepackage{times}
\usepackage{graphicx}
\usepackage{ulem}
\usepackage[colorlinks=true,citecolor=blue]{hyperref}
\newcommand{\beq}{\begin{equation}}
\newcommand{\eeq}{\end{equation}}

\begin{document} 

\title{Phase diagram and non-Abelian symmetry locking for fermionic
mixtures with unequal interactions}

\author{Joao C. Pinto Barros}
\affiliation{SISSA and INFN, Sezione di Trieste, Via Bonomea 265, I-34136 Trieste, Italy}

\author{Luca Lepori}
\affiliation{Dipartimento di Fisica e Astronomia, Universita di Padova, Via Marzolo 8, I-35131 Padova, Italy}

\author{Andrea Trombettoni}
\affiliation{CNR-IOM DEMOCRITOS Simulation Center, Via Bonomea 265, I-34136 Trieste, Italy}
\affiliation{SISSA and INFN, Sezione di Trieste, Via Bonomea 265, 
I-34136 Trieste, Italy}

\begin{abstract}
The occurrence of non-Abelian symmetry-locked states in ultracold 
fermionic mixtures with four components is investigated.    
We study the phase diagram 
in the presence of an attractive interaction between 
the species of two pairs of the mixture, and general (also repulsive) 
interactions between the species of each pair. This system 
is physically realized, e.g., in mixtures of two 
different earth-alkaline species, 
both of them with two hyperfine levels selectively populated. 
We find an extended region of the diagram
exhibiting a two-flavors superfluid symmetry-locking 
(TSFL) phase. This phase is present also for not too large repulsive 
intra-pair interactions and it is characterized 
by a global non-Abelian symmetry 
group obtained by locking together two independent
invariance groups of the corresponding normal state. 
Explicit estimates are reported for the mixture of the fermionic isotopes 
$^{171}\mathrm{Yb}$-$^{173}\mathrm{Yb}$, indicating that the TFSL phase 
can be achieved also without tuning the interactions 
between $\mathrm{Yb}$ atoms.
\end{abstract}

\maketitle 

\section{Introduction}
\label{sec:intro}

Ultracold atoms provide an ideal playground for the simulation 
of strongly interacting quantum systems \cite{bloch08},
mainly due to their high tunability and to the variety of the measurements
that can be performed on such systems. 
Two ingredients greatly increase the versatility of ultracold atomic systems: 
optical lattices \cite{LibroAnna} and gauge potentials \cite{dalibard11}. 
The wide class of phenomena that have been or may be studied 
using optical lattices
include Mott-superfluid transitions \cite{greiner02}, Anderson localization 
\cite{AndBEC-1,AndBEC-2}, Josephson physics \cite{cataliotti01} and 
Hubbard physics in fermionic mixtures
\cite{esslinger10}. 

Regarding gauge potentials, 
the internal degrees of freedom coupled with them 
are in general hyperfine levels of certain atoms \cite{gerbier10}.  
At the present time mostly static gauge potentials have 
been realized experimentally, however 
in last years proposals for dynamic gauge fields also appeared 
\cite{Dyn1,Dyn2,Dyn3,Dyn4,Dyn5} and recently the first experimental 
realization has been performed \cite{martinez2016}.

In strongly correlated condensed matter physics, 
gauge theories occur as effective models \cite{nagaosa99}, 
for instance for antiferromagnets \cite{Bal} and  
high-temperature superconductors \cite{SLHTC}. 
In this respect, the synthesis of Abelian and non-Abelian 
gauge potentials and fields, possibly on optical lattices 
\cite{jaksch03,osterloh05,aidelsburger12,jimenez12,hauke12}, 
is expected to boost in the next future the investigation of a larger 
set of interesting systems, phenomena and phases.

The realization of gauge potentials and fields 
points to the simulation of systems 
relevant for high energy physics, as QCD-like theories and strongly coupled 
field theories. The possibility of bringing, in an ultra-cold laboratory, 
paradigmatic models of high energy physics 
has been discussed intensively in recent years.
Notable proposals on this topics concern a variety of phenomena and models, 
including 
$2D$ \cite{Dirac2D,guineas,wu08,Juzeliunas,lim08,hou09,lee09,alba13} and 
$3D$ \cite{Lamata,LMT,RevCold} Weyl and Dirac fermions, 
Wilson fermions and axions \cite{toolbox}, 
neutrino oscillations \cite{neutosc}, 
extra dimensions \cite{extra}, symmetry-locked phases \cite{lepori2015}, 
curved spaces \cite{boada2015}, 
Schwinger pair production \cite{kasper2015} and CP$(N)$ models 
\cite{laflamme2016}. Theoretical proposals came along with 
experimental achievements, including the realization of 
Dirac fermions in honeycomb lattices \cite{Tarr} and of 
the topological Haldane model \cite{Haldane}.
Finally ultracold fermions probed useful to explore the unitary limit 
\cite{Zwerger}, sharing several common features  
with neutron stars physics \cite{dean03}: 
large interactions of the unitary limit could be used as tool to construct 
toy models for quark confinement, 
chiral symmetry breaking and string breaking.

A central concept for various areas of high energy physics
is symmetry-locking. This phenomenon occurs in the presence of a
phase (typically superfluid), characterized by a 
non vanishing vacuum expectation value, acting as an order parameter,
breaking part of the symmetry that occurs in the absence of it. 
In particular, because of this expectation value, two initially 
independent symmetry groups
are mixed in a residual symmetry subgroup.

Symmetry-locking results in a number of peculiar 
properties, especially when the locked groups are non-Abelian, 
for instances ordered structures as nets and crystals \cite{Raj, Manna} 
or vortices and monopoles with semi-integer fluxes, confining
non-Abelian modes 
\cite{Auzzi:2003fs,Shifman:2004dr,Hanany:2003hp,Eto:2005yh,Auzzi:2004if}.  
A remarkable example
of this phenomenon appears in the study of nuclear matter 
under extreme conditions,
as in the core of ultra-dense neutron stars \cite{alford1999}. 
There the locking interests the $SU(3)_c$ (local) color and the 
$SU(3)_f$ (global) flavor groups. 
Similarly the chiral symmetry breaking transition involves a locking of global
$SU(3)_L$ and right $SU(3)_R$ global flavor symmetries \cite{Raj,Manna}.

A step forward towards the study of symmetry-locked states 
was presented in \cite{lepori2015}, based on multi-component fermionic mixtures: 
there a proposal for the synthesis of a superfluid
phase locking two non-Abelian global symmetries 
has been presented. This state has been 
denoted as a two-flavour symmetry-locked (TSFL) state. 
In the analysis presented in \cite{lepori2015} it was 
considered a four component mixture with attractive interactions 
between the species of two pairs (denoted by $c$ and $f$) 
of the mixture (the interaction coefficient being denoted by $U_{cf}>0$) and 
attractive interactions between the species of the two pairs (respectively 
$U_c>0$ and $U_f>0$). With $U_c=U_f \equiv U$ and $U_{cf}>U$ 
the mixture hosts very peculiar phenomena belonging 
to the realm of high-energy physics,
as TSFL states, fractional vortices and non-Abelian modes confined on them 
\cite{lepori2015}. Beyond its intrinsic interest, this scheme 
represents a first step towards the simulation of phases 
involving the breaking of local (gauge) symmetries, as in the QCD 
framework.

Multi-component fermionic mixtures appear to be a natural playground
to simulate symmetry-locking. One notable example is given by multi-components $\mathrm{Yb}$ gases, 
that can be synthesized and controlled at the present time
\cite{Fallani}.  $\mathrm{Yb}$ atoms, as all the 
earth-alkaline atoms, 
have the peculiar property that their interactions do not depend 
on the hyperfine quantum number labelling the states of a certain multiplet. 
This fact allows to realize interacting systems, bosonic and fermionic, 
with non-Abelian $U(N)$ or $SU(N)$ symmetry \cite{alkaline}. 
In particular one could realize a four component mixture with attractive 
interactions between two pairs of species using 
a mixture of fermionic $^{171}\mathrm{Yb}$ and $^{173}\mathrm{Yb}$ atoms,
each species in two different hyperfine levels selectively populated and loaded on a cubic optical lattice. However, although the scattering length 
$a_{\mathrm{171-173}}$ between 
$^{171}\mathrm{Yb}$ and $^{173}\mathrm{Yb}$ atoms is negative and rather large 
($a_{\mathrm{171-173}}=-578a_0$, with $a_0$ the Bohr radius) 
resulting in $U_{cf}>0$, 
the scattering length $a_{\mathrm{171-171}}$ between between 
$^{171}\mathrm{Yb}$ atoms 
is small and negative ($a_{\mathrm{171-171}}=-3a_0$) giving $U_c\approx0$, and  
and the scattering length 
$a_{\mathrm{173-173}}$ between $^{173}\mathrm{Yb}$ atoms 
is {\it positive} and much larger than $a_{\mathrm{171-171}}$ 
($a_{\mathrm{173-173}}=+200a_0$) 
resulting in $U_{f}<0$, i.e. a repulsion \cite{taie2010}. 

The natural question arising from the discussion above is whether intra-pair repulsions  
(associated to $U_f<0$ in the example of the 
$^{171}\mathrm{Yb}$-$^{173}\mathrm{Yb}$ mixture) can destroy the TSFL phase 
induced by an inter-pair attraction. A related question is the determination 
of the phase diagram and the actual extension of the TSFL phase as the interactions
between the atoms of the considered four-component mixture are varied. 

In order to settle these questions, in  the present paper we explore the phase diagram of a four component 
mixture with attractive interactions between 
the species of two pairs of the mixture, and general (also repulsive) 
interactions between the species of the pairs, 
clarifying the ranges for the experimental parameters 
where a TSFL phase can occur.
By our study we conclude that, 
a TFSL phase could be synthesized in a close future, 
using already reachable values of the experimental parameters,
like the lattice widths. 
Notably this task can be achieved
just assuming the natural interactions of $^{171}\mathrm{Yb}$ and 
$^{173}\mathrm{Yb}$ atoms, 
without any external tuning. Indeed for instance the critical 
temperature required to enter in the superfluid TSFL phase
turns out of the same order of the ones presently reached. 
This results is particularly relevant
in the light of the known difficulty to tune interactions between 
earth alkaline atoms, as the $Yb$, without destructing their 
$U(N)$ invariance and avoiding important losses of atoms or warming
of the experimental set-ups.

\section{The Model}
\label{model}

We consider a four species fermionic mixture involving atoms in 
two different pairs of states (possibly pairs of hyperfine levels). 
For convenience we label the four degrees of freedom 
as $\sigma\in\left\{ r,g,u,d\right\}$ 
and distinguish between the species $\left\{ r,g \right\}$ 
in the first pair $c$ and the species 
$\left\{ u,d \right\} $ in the second pair $f$.
 
Even if the mechanism we are going to describe is 
independent on the space where the atoms are embedded, 
in the following the mixture will be considered loaded in 
a cubic optical lattice. A discussion of possible advantages of this choice 
will be presented in Section \ref{exper}. 
The system is described by an Hubbard-like Hamiltonian $H=H_{kin}+H_{int}$ 
\begin{equation}
\begin{array}{c}
H_{kin}=-t\sum\limits_{\left\langle i,j\right\rangle ,\sigma}c_{i\sigma}^{\dagger}c_{j\sigma} \, , \\
{}\\
H_{int}=- \sum\limits_{i,\sigma \sigma\prime}  U_{\sigma\sigma'} n_{i\sigma}n_{i\sigma^{\prime}}
\end{array}
\label{H}
\end{equation}
(with $(t>0)$). 
The matrix $U_{\sigma\sigma'}$ is
symmetric with vanishing diagonal elements (because of the Fermi statistics).  

We are interested in particular on a situation where the interactions 
between the multiplets $c$
and $f$ does not depend on the specific levels chosen in each pair. 
An experimental realization of the
this condition is performed by using earth-alkaline atoms.
For instance, a specific proposal relies on the use of the two hyperfine levels of $^{171}\mathrm{Yb}$ and of two suitably chosen levels in the $6$-multiplet
 of  $^{173}\mathrm{Yb}$. 
More details on this mixture will be given in Section \ref{exper}, see as well 
\cite{yip2011}.

The system \eqref{H} is therefore characterized by interactions 
labelled as 
$U_{rg} \equiv U_{c}$, $U_{ud} \equiv U_{f}$ and 
$U_{ru}=U_{rd}=U_{gu}=U_{gd} \equiv U_{cf}$. 
In the following we will refer to the interactions associated with $U_{c}$ and $U_{f}$ as "intra-pair" interactions and to the ones associated 
with $U_{cf}$ as "inter-pair" interactions.

Once the hoppings and the occupation numbers of the species are set equal 
in each multiplet, the system in the normal (Fermi liquid) state 
has a group symmetry $G=U\left(2\right)_{c}\times U\left(2\right)_{f}$
corresponding to {\it independent} rotations on the $c$ and $f$ degrees
of freedom respectively. On the contrary, as shown in \cite{lepori2015},
when superfluidity is induced, $G$ may undergo in general 
a spontaneous symmetry breaking into a smaller subgroup $H$. 
In particular when superfluidity occurs between the $c$ and the $f$ atoms, 
the following SSB pattern $G \to H$ takes place:
\begin{equation}
U(2)_c \times U(2)_f  \to U(2)_{c + f}.
\label{pattern1}
\end{equation}
This means that the superfluid phase 
has a residual non-Abelian invariance group 
$H = U(2)_{c + f}$ composed by a subset of the group of elements 
$({\cal U}_c, {\cal U}_f) = ({\cal U}_{c}, {\cal U}_{c}) = 
({\cal U}_{f}, {\cal U}_{f})$, where ${\cal U}_c$ and ${\cal U}_f$ 
belong to $U(2)_{c}$ and $U(2)_{f}$ respectively. Notably $H = U(2)_{c + f}$ 
involves at the same time $c$ and $f$ transformations, 
originally independent. 

The SSB at the basis of the symmetry-locking is explicit in the fact that the superfluid
is described by a gap matrix ${\bf \Delta}_{cf}$ transforming under $G$ as 
${\cal U}_{c}  \, {\bf \Delta}_{cf} \, {\cal U}_{f}^{-1}$, and left invariant
by the subgroup of transformations $H = U(2)_{c + f}$.
This mechanism is called symmetry-locking \cite{alford1999}.

\section{Mean field energy and consistency equations}
\label{MFeq}

In the present Section we consider the possible emergence of superfluid states, with various (numbers of) pairings in the system described by Eq. \eqref{H},
investigating more in general the superfluid BCS phases that can arise in it. 
We start the analysis by using a mean field approximation, and we 
present strong-coupling results in Section \ref{diagram}.

Omitting details, in the mean field approximation the energy ${\cal F}$ 
at zero temperature can be written as:
\begin{equation}
{\cal F}= \frac{1}{2} \, \underset{\vec{k}}{\sum} \, \hat{\psi}_{\vec{k}}^{\dagger} \, F_{\vec{k}} \, \hat{\psi}_{\vec{k}}+F_{c},
\label{eq:mean_field_matrix}
\end{equation}
where $\hat{\psi}_{\vec{k}}^{\dagger}=\left(c_{kr}\ldots c_{kd} \, , \, -c_{-kr}^{\dagger}\ldots -c_{-kd}^{\dagger}\right)$,
and $F_{\vec{k}}$ is  the $8\times8$ matrix: 
\begin{equation}
F_{\vec{k}}=  \, \left(\begin{array}{cc}
\xi_{\vec{k}, \{\sigma\}}   & 2 \Delta_{\sigma \sigma^{\prime}} \\
2 \Delta^{*}_{\sigma \sigma^{\prime}} & -\xi_{\vec{k}. \{\sigma\}}  
\end{array}\right),
\label{mat}
\eeq
In Eq. \eqref{mat} the factor $2$ in front of 
$\Delta_{\sigma \sigma^{\prime}}$ is due to the double sum in Eq. \eqref{H}. 
Moreover we set 
$$\xi_{\vec{k} \sigma}=\mathrm{Diag}\left(\varepsilon_{\vec{k}}-\tilde{\mu}_{\sigma}\right),$$ where
$$\varepsilon_{\vec{k}} = -2t \, \sum_{l=1}^3 \, \mathrm{cos} k_{\hat{l}}$$
and 
\begin{equation}
\tilde{\mu}_{\sigma}=\mu_{\sigma}+ \nu_{\sigma}U_{\sigma}+2 \nu_{\bar{\sigma}}U_{cf}
\label{tilde_mu}
\eeq 
are the chemical potentials shifted by the Hartree terms. 
In Eq. \eqref{tilde_mu} 
$\nu_{\sigma}$ denote the fillings and $\bar{\sigma}$ denotes 
the "opposite'' degree of freedom, 
so if $\sigma$ is a $c$ index then $\bar{\sigma}$ is an $f$ and
vice-versa. Notice that here we explicitly assume the balance between 
the two $c$ and the two $f$ species separately 
(this is the origin of the $2$ factor in front of $\nu_{\bar{\sigma}} \, U_{cf}$ 
in the expression above for $\tilde{\mu}_{\sigma}$).  

The constant $F_{c}$ in Eq. \eqref{eq:mean_field_matrix} is defined as follows:
\begin{equation}
F_{c}=\frac{1}{2}\underset{\vec{k},\sigma}{\sum}\xi_{\vec{k}\sigma}+ V\,\underset{\sigma\neq\sigma'}{\sum}U^{-1}_{\sigma\sigma'} |\Delta_{\sigma\sigma'}|^2,
\label{gssup}
\end{equation} 
$V$ being the number of the lattice sites,
$\left\langle c_{k\sigma}^{\dagger}c_{k\sigma'}\right\rangle =\delta_{\sigma\sigma'}n_{\sigma}$ and  
$\Delta_{\sigma\sigma'}\equiv -  \, V^{-1} \, U_{\sigma \sigma^{\prime}}  \, \underset{\vec{k}}{\sum}\left\langle c_{k\sigma}c_{-k\sigma'}\right\rangle $, assumed real.  
Moreover  
$\mu_{r}=\mu_{g}\equiv\mu_{c}$ and $\mu_{u}=\mu_{d}\equiv\mu_{f}$.

The problem to describe superfluid phases of the Hamiltonian in Eq. \eqref{H} 
is then reduced, at the mean field level, to the diagonalization of 
$F_{\vec{k}}$ and to the subsequent determination of of $\Delta_{\sigma\sigma'}$
and $\tilde{\mu}_\sigma$ by the solution of self-consistent equations. 
Of course, if more solutions are found one has to choose the one having 
the smaller energy.

The energy of the system can be found diagonalizing the 
matrix $F_{\vec{k}}$ and obtaining 
its eigenvalues $\lambda_{\vec{k} , \alpha}$, with 
$\alpha = 1, \dots,8$, divided in two sets with opposite sign. 
Putting the resulting diagonal form of $\cal{F}$ in normal order, 
all the eigenvalues are defined positive; in this way the constant term
$F_c$ is shifted as 
$F_{c}\rightarrow F_{c}-\underset{\vec{k},\alpha}{\sum} \, \frac{\lambda^{(+)}_{\vec{k} , \alpha} }{2}$, where $\lambda^{(+)}_{\vec{k} , \alpha}$ denote the four positive eigenvalues of $F_{\vec{k}}$. 

The ground-state energy is found to be
\begin{equation}
F_c =\frac{1}{2}\underset{\vec{k}}{\sum} \Bigg(\underset{\sigma}{\sum} \xi_{\vec{k}\sigma} - \underset{\alpha}{\sum} \lambda^{(+)}_{\vec{k}\alpha}\Bigg) +V \,\underset{\sigma\sigma'}{\sum}U^{-1}_{\sigma\sigma'} \left|\Delta_{\sigma\sigma'}\right|^{2}.
\label{gsen}
\end{equation}

The self-consistent equations for $\Delta_{\sigma, \sigma^{\prime}} $ 
and the shifted chemical potentials $\tilde{\mu}_{\sigma}$ 
can be now obtained from the conditions:
\beq
\left\{\begin{array}{c}
\frac{\partial {F_c}}{\partial \Delta_{\sigma, \sigma^{\prime}} }= 0 \\
{}\\
\frac{\partial \big({ F_c} \, + \, \tilde{\mu}_{\sigma} n_{\sigma} \big)}{\partial \tilde{\mu}_{\sigma}} = 0.
\end{array} \right.
\label{gapeq}
\eeq 

Several solutions of the Eqs. \eqref{gapeq} are possible in general. 
For this reason to fix the correct phase for every point $\big(\frac{U_c}{t},\frac{U_f}{t}, \frac{U_{cf}}{t} \big)$ of the diagram 
one has to find the lowest-energy solution.

We distinguish the various solutions as follows:
\begin{itemize}
\item Normal: no superfluid pairing exist between any degree of freedom. 
That means $\Delta_{\alpha\beta}=0$ for any pair $(\alpha,\beta$).
\item non-TSFL (NTSFL): intra-pair pairings occur but no inter-pair ones: 
$\left|\Delta_{c_{1}c_{2}}\right|^{2}+\left|\Delta_{f_{1}f_{2}}\right|^{2} \neq 0$ and  ${\bf \Delta}_{cf}=0$. In this case the two non-trivial Bogoliubov energies 
entering Eq. \eqref{gsen} read 
$\lambda^{(+, c)}_{\vec{k}\alpha} = \sqrt{\xi_{\vec{k}}^{2}+
\left|\Delta_{c_{1}c_{2}}\right|^{2}}$ and 
$ \lambda^{(+,f)}_{\vec{k}\alpha} = \sqrt{\xi_{\vec{k}}^{2}+\left|\Delta_{f_{1}f_{2}}\right|^{2}}$.
\item TSFL: inter-pair pairings occur but no intra-pair ones: 
$\left|\Delta_{c_{1}c_{2}}\right|^{2}+\left|\Delta_{f_{1}f_{2}}\right|^{2}=0$ and ${\bf \Delta}_{cf}\neq0$. 
In this case the two non-trivial Bogoliubov energies entering 
Eq. \eqref{gsen} read $\lambda^{(+)}_{\vec{k}\alpha} = \sqrt{\xi_{\vec{k}}^{2}+|\Delta_{cf}|^{2}}$ with $\Delta_{cf}=\frac{1}{2} \, \mathrm{Tr}{\bf \Delta}_{cf}$, being ${\bf \Delta}_{cf}$ the matrix of the inter-pair pairings.
\end{itemize}

Solving numerically Eqs. \eqref{gapeq}, it turns out that whenever in the presence of an attraction term between the species ($U_{cf}>0$), apart from the 
normal state solution, a solution with non-zero pairing 
$\Delta_{\sigma\sigma'}$ and energy lower than the normal state always exists. 
This result assures the presence of a superfluid state, also in presence 
of intra-pair repulsion. Of course this is a mean field result, expected 
not to be correct for large intra-pair repulsions: a strong-coupling analysis 
of such case is presented in Section \ref{diagram}.

The obtained superfluid BCS solutions are always of the TFSL or NTFSL types, 
in other words no solution with both $\left|\Delta_{c_{1}c_{2}}\right|^{2}+\left|\Delta_{f_{1}f_{2}}\right|^{2} \neq 0$ and ${\bf \Delta}_{cf}\neq0$ occurs. 
We observe that setting $n_c = n_f$ 
for all the three mentioned types of solutions, the shifted chemical 
potential $\tilde{\mu}_c$ and $\tilde{\mu}_f$ turn out equal, 
in spite of the intra-pair interactions $U_c$ and $U_f$, different in general. 
In particular, they depend only on $n_c$ and $n_f$ themselves. 
This means that, at least at the mean field level, 
these interactions do {\it not} determine any effective unbalance between 
the $c$ and $f$ species. This fact is expected to remain 
at least approximatively true in the presence of 
a trapping potential, since this potential acts, 
in local density approximation, as a space-dependent correction 
to the chemical potentials $\mu_{c, f}$ at the center of the trap 
\cite{pethick}, not to the shifted potentials $\tilde{\mu}_{c, f}$. 
This appears particularly relevant since it is 
known (see \cite{pethick} and references therein) that 
generally an unbalance in the normal state can spoil the possible 
emergence of superfluid states, or at least to modify the critical 
interaction strength and the critical temperature.

For the case $n_c = n_f \equiv n$, it is true that
$\xi_{\vec{k}, \sigma} \equiv \xi_{\vec{k}}$ and it is possible to 
recast the self-consistency Eqs. \eqref{gapeq} in a BCS-like form:
\begin{equation}
1= \frac{U_{c , f}}{V} \underset{\vec{k}}{\sum}\frac{1}{\sqrt{\xi_{\vec{k}}^{2}+ 4\left|\Delta_{c , f}\right|^{2}}}, \quad \Delta_{cf}=0,\mathrm{\hspace{0.7cm}NTSFL}
\label{self1}
\end{equation}
or 
\begin{equation}
1= \frac{U_{cf}}{V}\underset{\vec{k}}{\sum}\frac{1}{\sqrt{\xi_{\vec{k}}^{2}+ 4\left|\Delta_{cf}\right|^{2}}}, \quad \Delta_{c , f}=0,\mathrm{\hspace{0.7cm}TSFL}
\label{self2}
\end{equation}
and
\begin{equation}
n_{\theta} = \frac{1}{V} \, \underset{\vec{k}}{\sum} \Bigg(1-\frac{\xi_{\vec{k}}}{\sqrt{\xi_{\vec{k}}^{2}+ 4 \left|\Delta_{\theta}\right|^{2}}} \Bigg).
\label{self3}
\end{equation}
For sake of brevity, in the last equation $\Delta_{\theta}$ 
is meant to include both $\Delta_{c f}$ and $\Delta_{c},\Delta_f$, 
corresponding to both the cases TFSL and NTFSL. 
Notice that Eqs. \eqref{self1}-\eqref{self3} reproduce exactly the standard 
BCS self-consistency equations, as one should expect: indeed 
the different numerical factors in Eqs. \eqref{self1}-\eqref{self3} are 
due to the different definitions for $U_{c}$, $U_f$, $U_{cf}$ and for the 
corresponding gap parameters used here.

\section{The phase diagram}
\label{diagram}

In this Section we use Eqs. \eqref{gsen}-\eqref{gapeq}) to investigate 
the phase diagram of the Hamiltonian \ref{H} as a function of 
the external parameters $t$, $U_{c}$ $U_f$ and $U_{cf}$. 
In particular, we numerically solve Eqs. \eqref{gapeq}) 
for a cubic lattice having $20^3$ sites (checking that the phase diagram 
is not affected by finite size effects),  and we compare the energies 
of the obtained solutions to determine the mean field phase diagram. Later 
on the text we discuss limitations of the mean field findings and 
an alternative approach to study the case of large intra-pair 
repulsive interaction.

\begin{figure}
\begin{centering}
\includegraphics[scale=0.35]{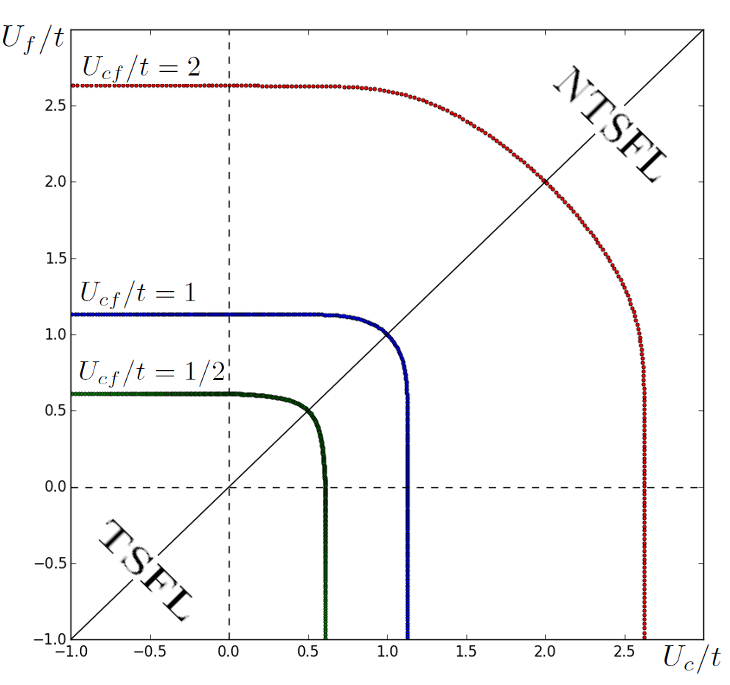}
\end{centering}
\caption{Phase diagram at half filling for $U_{cf}/t=\{1/2,1, 2\}$. 
Inside the curves (at smaller values of $U_\sigma$) the TSFL phase occurs, 
while outside one has the NTSFL phase. As $U_{cf}/t$ increases, 
the zone of the TSFL phase becomes larger.}
\label{fig:phased1}
\end{figure}

\begin{figure}
\begin{centering}
\includegraphics[scale=0.35]{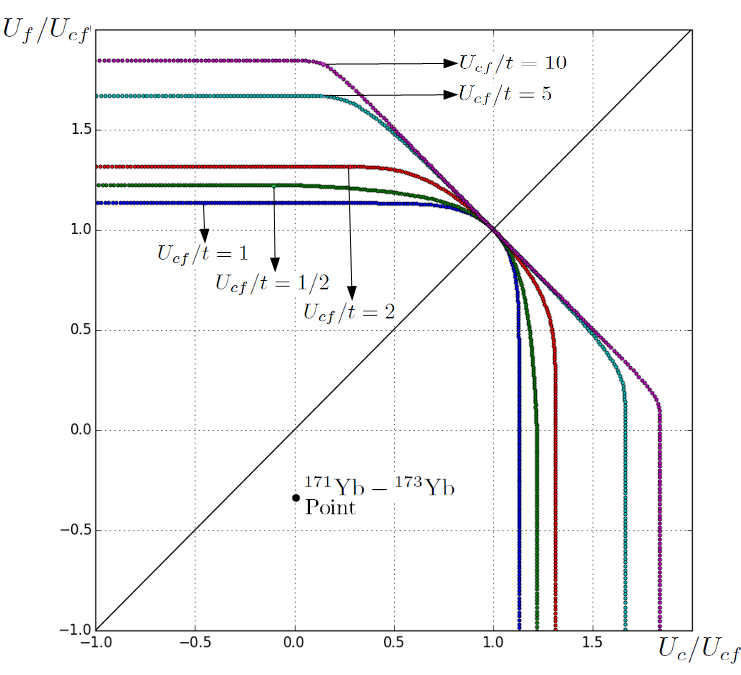}
\end{centering}
\caption{Phase diagram in units of $U_{cf}$ at half filling. 
The point $U_c=U_f=U_{cf}$ is a transition point between the phases TFSL 
and NTFSL, irrespectively of the value for $t$. It is also depicted 
the point representing the natural interactions of the mixture 
$^{171}\mathrm{\mathrm{Yb}}$-$^{173}\mathrm{\mathrm{Yb}}$ \cite{yip2011}. 
The corresponding estimates for this point are performed in Sec. \ref{exper}.}
\label{fig:phased2}
\end{figure}

\subsection{Attractive $U_{c}$, $U_f$}

The results presented in Fig. \ref{fig:phased1} refer to the the 
half-filling case \big($n_{\sigma} = \frac{1}{2}$, corresponding to 
$n_c = n_f \equiv n = 1$\big) and different values of the ratio 
$U_{cf}/t$ and $U_{c}/t$, $U_f/t$. In this case we always find 
$\tilde{\mu}_{\sigma} = 0$, as required by particle-hole symmetry 
(see e.g. \cite{fradkin2013}). 

For each fixed value of $U_{cf}/t >0$ (attractive regime)
a colored curve is drawn, separating the TSFL phase inside of it
from the NTSFL phase outside. We see that, 
as we increase the value of $U_{cf}/t$, higher values of attractive intra-pair 
couplings $U_{c}/t$, $U_f/t$ are required to break the TFSL phase 
in favour of the NTSFL one. At variance the normal state is never favored 
over both the superfluid states, even when one of or both the 
intra-pair interactions are repulsive and not small in comparison 
with the attractive ones. In this case the mean field approach is expected 
 not to be reliable and,  as we will see in the next Subsection, 
antiferromagnetic states can be instead favoured.

In Fig. \ref{fig:phased2} the curves of Fig. \ref{fig:phased1} 
are rescaled by their values of $U_{cf}/t$: in this way they all meet in the 
point $U_{c}=U_{f}=U_{cf}$. In this point all 
the different Hamiltonians have a $U\left(4\right)$ symmetry and 
the two phases TFSL and NTFSL can be mapped onto each other, 
signaling a transition point between the two phases, in agreement with 
\cite{lepori2015}.  

\begin{figure}
\begin{centering}
\includegraphics[scale=0.35]{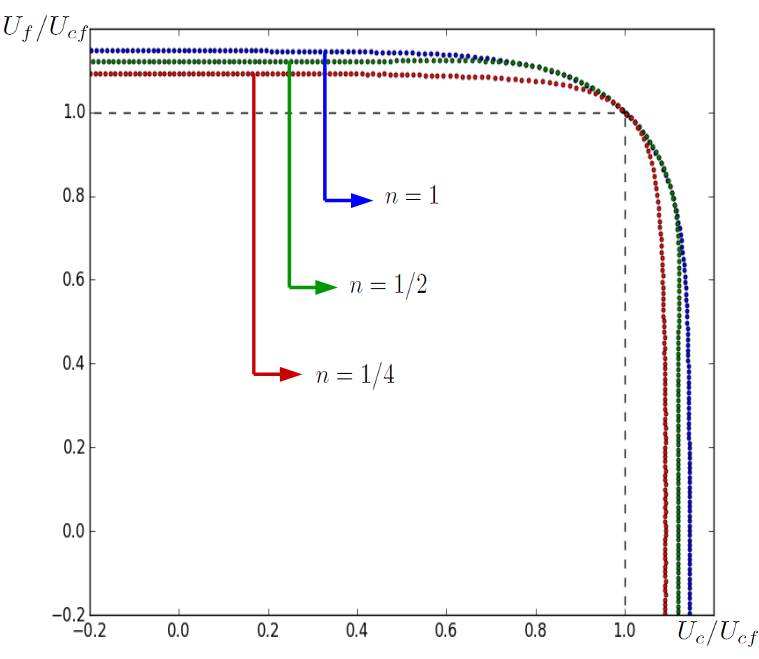}
\par\end{centering}
\caption{Phase diagrams for $U_{cf}/t =1$ and for different fillings: 
$n=1$ (blue), $n=1/2$ (green) and $n=1/4$ (red). They appear qualitatively 
very similar indicating that the filling does not play 
a fundamental role.}
\label{awayhalf1}
\end{figure}

\begin{figure}
\begin{centering}
\includegraphics[scale=0.35]{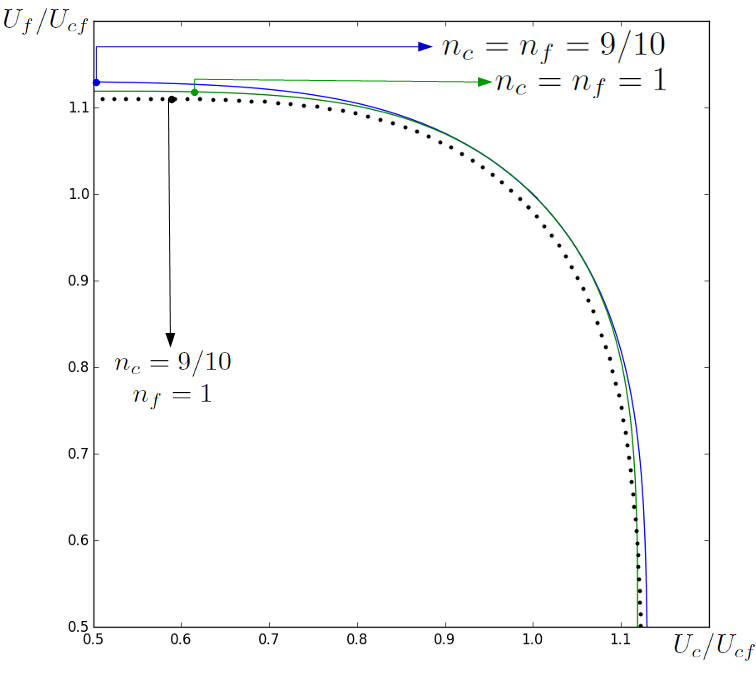}
\par\end{centering}
\caption{Phase diagram in the presence of a small unbalance 
between the  populations $n_f-n_c = 0.1$ and $U_{cf}=t$. 
The result is qualitatively very similar to the balanced cases 
(see also Figs. \ref{fig:phased2} and \ref{awayhalf1}).}
\label{awayhalf2}
\end{figure}

The black point in  Fig. \ref{fig:phased2} represents 
the case of the mixture composed by $^{171}\mathrm{\mathrm{Yb}}$ and 
$^{173}\mathrm{\mathrm{Yb}}$, where natural interactions between 
these isotopes are also assumed. This mixture, mentioned in 
Section \ref{sec:intro}, will be discussed in detail 
in Section \ref{exper}. 
Here we notice only that the point lies well inside the TSFL zone.

The phase diagram shown in Fig. \ref{fig:phased2} is not a 
consequence of the hypothesis of balanced mixture. 
Indeed in Fig. \ref{awayhalf1} we plot the same phase diagram for 
different fillings (but still equal for the four $\sigma$ species), 
finding qualitative agreements with small quantitative differences. 
Similarly, in Fig. \ref{awayhalf2} the case where the pairs $c$ and $f$ 
have fillings differing by ten percent is reported. 
Again we see that the imbalance in the populations does not produce 
significative differences on the results. We stress that, 
although an imbalance in the number of particles is generally known 
able to spoil the appearance of superfluid states \cite{pethick}, 
in the present case the reliability of our results is 
guaranteed by the absence of other non-trivial solutions for the 
Eqs. \eqref{gapeq} (see for comparison, e.g., \cite{mannarelli2006}) 
and by the direct comparison between the energies of the normal states 
and the one of the BCS-like superfluid solutions. 

\subsection{Repulsive $U_{c}$, $U_f$}
\label{rep}

When $U_{c}$, $U_f$ assume negative values and repulsive 
intra-pair interactions appear in the Hamiltonian of Eq. \eqref{H}, 
the formation of intra-pair pairs start to become suppressed. However 
the normal state is never favored in the mean field 
approximation as shown in Figs. \ref{fig:phased1}-\ref{awayhalf2}. 

If it is reasonable that for small intra-pair 
repulsions the TSFL is favoured, 
for large enough values of $U_{c}/t$, $U_f/t$ and 
$U_{c}/U_{cf}$, $U_f/U_{cf}$ this superfluid phase 
is expected to eventually disappear, replaced by insulator 
phases with a magnetic-like order. The latter regime is qualitatively 
described  in the strong coupling limit $U_{c}/t$, $U_f/t$ by spin Hamiltonians, similarly to the Heisenberg model for a two species repulsive mixtures 
at half filling (see, e.g., \cite{fradkin2013}).

In the strong-coupling limit two cases are explicitly considered here: 
$a)$ $|U_{c}|/t, |U_f|/t \gg1$,  $b)$ $|U_{c}|/t \ll1$ and $|U_f|/t \gg1$. 
Notice that in both cases the further condition 
$|U_{c}/U_{cf}|, |U_f/U_{cf}| \gg 1$ is implicitly assumed. 

In the first case the strong coupling Hamiltonian reads 
(details of the derivation are in the Appendix \ref{appb}):
\beq
\hat{H}^{cf}_{eff}=\frac{t^{2}}{4}\underset{\left\langle i,j\right\rangle }{\sum}\left(\frac{1}{\left|U_{c}\right|}\vec{C}_{i}\cdotp\vec{C}_{j}+\frac{1}{\left|U_{f}\right|}\vec{F}_{i} \cdotp\vec{F}_{j}\right)- E_{GS}^{cf},
\label{eqeff1}
\eeq
where $\vec{C}$ and $\vec{F}$ are effective spin variables
defined by $\vec{S}_{i}=\underset{\sigma\sigma^{\prime}}{\sum}c_{i\sigma}^{\dagger} \vec{\tau}_{\sigma\sigma^{\prime}} c_{i\sigma^{\prime}}$ 
($\vec{\tau}$ denoting the Pauli matrices) and $E_{GS}^{cf}$ is the ground-state energy given by
\beq
E_{GS}^{cf} = -NU_{cf} -\frac{zNt^{2}}{4}\left(\frac{1}{\left|U_{c}\right|}+\frac{1}{\left|U_{f}\right|}\right),
\label{eg1}
\eeq
where $N = 2V$ is the total number of atoms of each pair. 
The Hamiltonian in Eq. \eqref{eqeff1} corresponds to 
two decoupled Heisenberg models.

The case $b)$ is of interest for the $\mathrm{Yb}$ discussed in the 
next Section, in the perspective of a possible experimental 
realization for the TFSL mechanism. Here the ground-state energy 
is found in the limit $U_c/t \to 0$ 
(see details in Appendix \ref{appc}):
\begin{equation}
E_{GS}^c = 2 E_{GS}^{NS} + \Delta E=  2 E_{GS}^{NS} - N\left(\frac{U_{c}}{4} + \frac{z \, t^{2}}{4\left|U_{f}\right|} \right),
\label{eg2}
\end{equation}
where $E_{GS}^{NS}$ is the energy of a single $c$ component in the normal state. 
Indeed the energy in Eq. \eqref{eg2} is proper of a system of free fermions 
$c$ on a antiferromagnetic background describing the dynamics of the $f$ 
fermions and described by a spin Hamiltonian similar to the one in 
Eq. \eqref{eqeff1}.

The regions of the phase diagram where both the 
TFSL and NTSFL superfluid phases 
occur can be bounded comparing their ground-state energies
with the energies of the antiferromagnetic phases in Eqs. \eqref{eg1} and 
\eqref{eg2}.

\begin{figure}[h]
\begin{centering}
\includegraphics[scale=0.35]{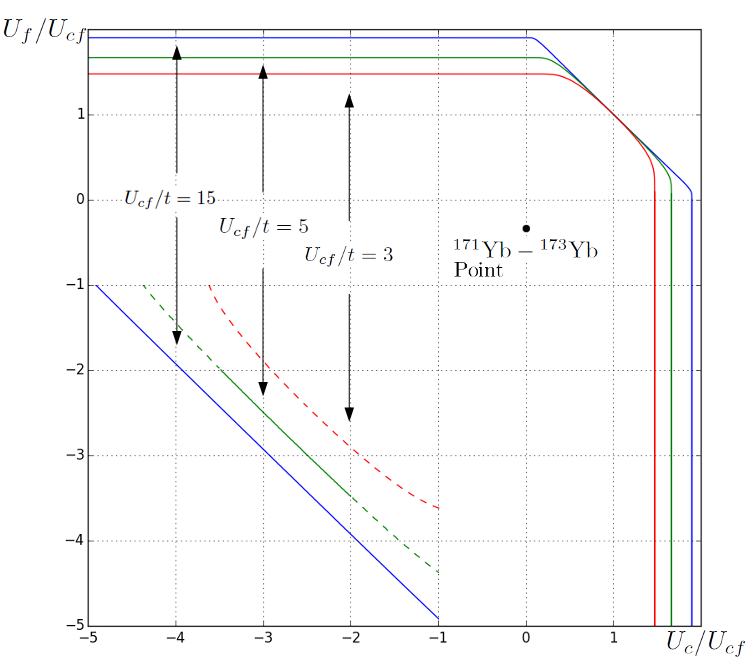}
\end{centering}
\caption{Phase diagram, containing the natural point for the Ytterbium mixture, for the cases $U_{cf}=3t$ (red), $U_{cf}=5t$ (green) and $U_{cf}=15t$ (blue). 
The oblique lines
bounding the superfluid phases are obtained by the strong coupling approach 
leading to Eqs. \eqref{eg1} and \eqref{eg2}.  
The transition from solid lines to dashed lines signals where this approach 
is not reliable any longer because it does not hold $|U_{c}/t|, |U_f/t| \gg 1$.}
\label{phaseant}
\end{figure}

Postponing the details for the case $b)$ to the Section \ref{exper}, 
we presente the results of this calculation for the case $a)$ in 
Fig. \ref{phaseant}. There the oblique lines 
represent a set of points where, according to 
the energy criterium mentioned above, 
the insulator states become favorable over the superfluid phases. 
Notice that increasing the depth $V_0$ results in a increase of the 
area of the TSFL phase, compared with the insulator one.
 
The calculation leading to Eqs. \eqref{eg1} and \eqref{eg2} 
is  perturbative in $t/U_{\sigma}$, therefore the comparison 
between the energies in the same equations and the ones 
for the superfluid states is reliable only $t/U_{\sigma} \ll1$. 
For this reason a dashed line, instead of a solid one, 
is drawn in Fig. \ref{phaseant} where the condition 
$|t/U_{\sigma}|>10^{-1}$ (a threshold conventionally chosen) starts to hold, 
so that  the strong-coupling approach is no longer expected 
to be fully reliable. From the figure we see that for $U_{cf}/t=3$ 
the transition line can never be located perturbatively, 
while for $U_{cf}/t=15$ the converse is true. As an intermediate 
example $U_{cf}/t=5$ exhibits both a zone where perturbation 
theory can be assumed valid and other ones where it cannot. 

\section{Experimental feasibility and limits}
\label{exper}

As mentioned in the Introduction, a possible experimental 
realization of the system investigated in the last Section 
is provided by a mixture of $^{171}\mathrm{\mathrm{Yb}}$
and $^{173}\mathrm{Yb}$.  
The first isotope has a $1/2$ hyperfine multiplet 
while the second one has $5/2$ hyperfine degeneracy. For the latter atomic 
species only two levels
could be selectively populated.
The mixture obtained in this way exhibits natural interactions 
characterized as follows: using conventionally the label $c$ 
for the hyperfine levels of $^{171}\mathrm{Yb}$ 
and the label $f$ for the ones of $^{173}\mathrm{Yb}$, the
scattering lengths are $a_{c}=-3a_{0}$, $a_{f}=200a_{0}$ and $a_{cf}=-578a_{0}$, 
where $a_{0}$ is the Bohr radius (see e.g. \cite{taie2010,yip2011}). 
As in all the earth-alkaline atoms, 
the tunability of these interactions is very difficult 
using the magnetic Feshbach resonance, because of the
negligible magnetic moment of such atoms. Moreover, in the recent literature 
this problem revealed challenging also using alternative techniques, 
due to important atomic losses and without spoiling
their characteristic $U(N)$ invariance ($N$ 
denoting here the number of hyperfine levels of the considered atomic species). 
For details on this subject see \cite{pagano2015} and references therein.
This problem can prevent the realization of certain phases and the exploration 
of the full phase diagram. For our purposes the question is then 
if without tuning the interaction the TSFL superfluid phase is realized or not.

For the considered earth-alkaline mixture loaded on a cubic lattice, 
the hopping parameters, in principle different, are given by: 
\beq
\begin{gathered}
t_{\alpha}=-\int d^{3}\vec{r} \,
\Bigg(\frac{\hbar^{2}}{2m_{\alpha}}\nabla\phi_{\alpha\vec{r}^\prime}\left(\vec{r}\right)\cdot\nabla\phi_{\alpha\vec{r}^{\prime\prime}}\left(\vec{r}\right)
+\\
+ \,  \phi_{\alpha\vec{r}^{\prime}}\left(\vec{r}\right)V_{\mathrm{ext}}\left(\vec{r}\right)\phi_{\alpha\vec{r}^{\prime\prime}}\left(\vec{r}\right) \Bigg).
\end{gathered}
\label{hops}
\end{equation}
The expressions for the interaction parameters $U_c, U_f, U_{cf}$  in the form of $U_{\alpha\beta}$ for $\alpha\neq\beta\in\left\{ r,g,u,d\right\}$ 
are [notice the minus sign in \eqref{H}]:
\beq
U_{\alpha\beta} = - \frac{\pi\hbar^{2} a_{\alpha\beta}}{m_{\alpha\beta}}\int d^{3}\vec{r}\left|\phi_{\alpha\vec{r}^{\prime}}\left(\vec{r}\right)\right|^{2}\left|\phi_{\beta\vec{r}^{\prime}}\left(\vec{r}\right)\right|^{2}.
\label{inter}
\end{equation}

In Eqs. \eqref{hops} and \eqref{inter}, 
 $\phi_{\{\alpha,\beta\}\vec{r}^{ \, \prime}}\left(\vec{r}\right)$ 
are the Wannier functions describing the localization on a given 
lattice site $\vec{r}^{ \, \prime}$ (these labels are suppressed in the following 
for sake of brevity), $\vec{r}$ is the distance from a chosen site, 
and $m_{\alpha\beta} = \frac{ m_\alpha \, m_\beta}{m_\alpha + m_\beta}$.
A simple variational estimate for the Wannier functions, 
which results in an estimate for the parameters in Eqs. \ref{hops} 
and \ref{inter}, is discussed in Appendix 
\ref{appA}. 

The tight binding regime for the $\mathrm{Yb}$ 
is achieved for $V_0 \gtrsim 2-3 E_{R_c}$ where 
$E_{Rc}=\frac{\hbar^{2}k_0^{2}}{2 m}$ is the recoil energy, 
$k_0$ is the wave vector of the laser producing the optical lattice and 
$m$  is chosen conventionally to be the mass of the 
$^{171}\mathrm{Yb}$ isotope. We consider $V_0$ up to $\approx 15 E_{R_c}$, 
where the tunneling coefficients are very small and tunneling dynamics 
effectively suppressed. 
Assuming this interval for the ratio $V_0/E_{R_c}$ and Eqs. \eqref{hops} 
and \eqref{inter} with their optimized Wannier wavefunctions, 
the regions on the diagram $U_{c}/U_{cf}$, $U_f/U_{cf}$ associated with 
the considered $\mathrm{Yb}$ mixture with natural interactions 
can be calculated. 

\begin{figure}
\begin{centering}
\includegraphics[scale=0.33]{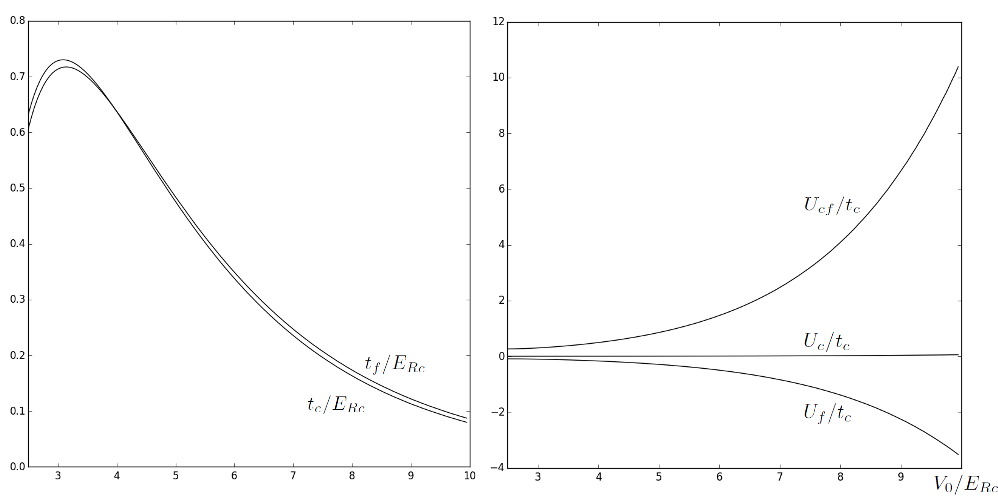}
\par\end{centering}
\protect\caption{Parameters of the Hamiltonian in Eq. \eqref{H} 
as a function of the depth of the optical lattice 
potential $\tilde{V}_{0}=V_{0}/E_{Rc}$.
Left panel: hopping parameters $t_{c}/E_{Rc}$ and $t_{f}/E_{Rc}$. 
Right panel: rescaled interaction parameters $U_{c,f}/t_c$ and $U_{cf}/t_c$.}
\label{param}
\end{figure}

In Fig. \ref{param} we report on the left panel the hopping coefficients
for different rescale depths $\tilde{V}_{0}=V_{0}/E_{Rc}$. 
We see in the left panel that, also considering the small 
difference in mass between the two isotopes, 
it always holds $\Delta t/t \lesssim 10^{-1}$ so that the previous 
assumption $t_c=t_f  \equiv t$ (however not strictly required for 
the TFSL mechanism) is reasonable. 
On the right panel of the same figure we report 
the variation of $U_{\alpha , \beta}/t$, again as functions 
of $\tilde{V}_{0}=V_{0}/E_{Rc}$. 
In the same way, the region in the diagram $U_{c,f}/U_{cf}$ 
associated with the $\mathrm{Yb}$ mixture can be also calculated. 
More details on the calculation are given in Appendix \ref{appA}. 

We observe that, once the intra-pair interactions 
are written in the form $U_{c,f}/U_{cf}$, the dependence on 
the amplitude $V_0$ effectively drops out, such that only 
the relative value of $U_{cf}/t$ changes significantly and the 
obtained region resembles a single point. 
This is the reason why we can speak about just a "natural point' 
in the diagrams of Figs. \ref{fig:phased2} and \ref{phaseant}. 
This point is given approximately by the coordinates 
$U_c/U_{cf} \approx 0.01$ and $U_f/U_{cf} \approx -0.34$,
also very close to the point estimated using the approximation 
$U_{\alpha}/U_{cf}\simeq a_{\alpha}/a_{cf}$ valid in the continuous space limit. 

Importantly the natural point falls well inside the TSFL regime, 
see Figs. \ref{fig:phased2} and \ref{phaseant}. In particular, along the 
line $U_c/t=0$ (case $b$ in Section \ref{diagram}), where 
the point almost lies, an estimate for the appearance 
of the antiferromagnetic regime can be done  comparing the energies 
in Eqs. \eqref{gssup} and  \eqref{eg2}. 
As a result,  the transition is located by the strong coupling 
approach at the values  $U_c/U_{cf} = -3.97$ for $U_{cf}/t=3$, 
$U_c/U_{cf} \approx -4.9 $ for $U_{cf}/t=5$ and $U_c/U_{cf} \approx - 5.6$ 
for $U_{cf}/t=15$, in all the three cases far 
from the natural point of the $\mathrm{Yb}$. 
In this way, our findings indicate that the TSFL phase 
can be observed in the zero temperature limit in experiments 
with $\mathrm{Yb}$ mixtures, assuming natural interactions 
and realistic values for the depth of the lattice potential.

Despite of the zero-temperature results reported, the 
TSFL phase could be still unreachable in the presence of a 
too low critical temperature (at fixed interactions) required for 
its emergence, in comparison with the ones currently realizable.  
This point is particular important in the light of the 
mentioned difficulty to tune the interactions in earth-alkaline atoms. 
We can make an estimate of the critical temperature for the $\mathrm{Yb}$ 
mixture. Proceeding as for the two-component attractive Hubbard model 
\cite{iazzi2012}, 
in the present case we refer to the case of isotropic
hoppings $t$ ($t_{\perp}=t_{\parallel} =t$ in the notation of \cite{iazzi2012}) 
and to the half filling case. Moreover  $\tilde{\mu}_c = \tilde{\mu}_f$, 
as we found in Section \ref{MFeq}.

For our model on a cube lattice 
the total bandwidth is $D=12t$. If we consider for instance $V_{0}=5E_{Rc}$, 
we obtain $2U_{cf} \approx0.3D$, which results in $T_c K_{B}/ D\approx0.05$. 
Using these values and considering a lattice spacing of $a=0.5 \, \mu m$,
the critical temperature turns out $T_{c}\approx15\ \mathrm{nK}$. 
In terms of the Fermi temperature this amounts to obtain $T_{c}/T_{F}\approx0.1$.
This value  is reasonably close to the ones achievable in current experiments 
\cite{valtolina15}, suggesting that 
the critical temperature assuming the natural interaction 
is reachable with current-day experiments and the TSFL phase 
could be achieved.

The lattice ratio $T_{c}/T_{F}\approx0.1$ can be compared with the typical 
one for experiments in the continuous space, finding that 
apparently on the lattice $T_c/T_F$ is sensibly larger. Indeed a very simple estimate can be 
done using the results \cite{varenna2006} for a two-component mixture 
(as it is effectively the TSFL phase). Considering a 
number of loaded atoms $N \approx 10^4$ and a system size 
$\ell \sim 10 \mu m$, one obtains $T_c/T_F$ smaller 
than $0.01$. 
This value is far from the presently achievable ones, 
differently from the lattice case. Summing up, 
the present analysis suggests that, for the task to synthesize a 
TFSL phase in $\mathrm{Yb}$ mixtures, 
the use of a (cubic) lattice can be advantageous.

\section{Conclusions}

In this paper we investigated the possible emergence of a 
non-Abelian two-flavor
locking (TSFL) superfluid phase in ultracold Fermi 
mixtures with four components and unequal interactions.
More in detail, using mean field and strong coupling results, we explored 
the phase diagram of this mixture loaded in 
a cubic lattice, finding 
for which ranges of the interactions and of the 
lattice width the system exhibits a TSFL phase. 

These ranges are found to have an extended overlap with 
the ones realizable in current experiments. 
In particular, as detailed in the text, the proposed set-up and phase are found 
to be realistic and realizable using a mixture
of $^{171}\mathrm{Yb}$ and $^{173}\mathrm{Yb}$. The phase diagram has been 
studied and the point in the phase diagram associate to the 
natural (not tuned) interactions 
between these atomic species determined. 
The critical temperature required for the appearance of the TSFL superfluid
has been found comparable with the ones currently achievable.
The latter ingredient is central for a possible experiment 
aiming to realize the TFSL phase,
especially due to the known difficulty to tune interactions
in earth-alkaline atomic gases, without spoiling their peculiar 
$U(N)$ invariance.

We finally observe that for our results it is crucial that relative large 
intra-pair repulsions do not destroy the superfluid states. Different is expected to be the 
case where non-local repulsive interactions are present, whose effects can be considered
an interesting subject of future work.

\section*{Acknowledgements}
The authors are pleased to thank F. Becca, A. Celi, 
L. Fallani, M. Mannarelli, F. Minardi, L. Salasnich 
and W. Vinci for useful discussions.

\appendix

\section{$c$ and $f$ strongly coupled limit\label{sec:CFlimit}}
\label{appb}

In this Appendix we present details of 
the perturbative calculation for the strongly coupled limit 
in the presence of repulsive intra-pair interactions, 
leading to Eq. \eqref{eqeff1}
in the main text. We consider half filling.

The described physical situation corresponds to consider 
the Hamiltonian $H_0+ H_1$ 
\begin{equation}
H_{0}= 2\underset{i}{\sum}\left(|U_{c}| n_{ir}n_{ig}+ |U_{f}| n_{iu}n_{id}\right)-  2 |U_{cf}| \underset{i,c,f}{\sum}n_{ic}n_{if},
\end{equation}
\begin{equation}
H_{1}= -t \underset{\left\langle i,j\right\rangle ,\sigma}{\sum}c_{i\sigma}^{\dagger}c_{j\sigma},
\end{equation}
and perform perturbation theory in the parameters 
$\varepsilon_{c},\varepsilon_{f} \ll 1$ with 
$\varepsilon_{c}=t/|U_{c} |, \varepsilon_{f}=t/|U_{f} |$. 
We assume $\varepsilon_{c} = \varepsilon_{f} =\varepsilon$. 
The ground-states of $H_{0}$, with energies $E=- 2V |U_{cf}| = - N |U_{cf}|$,
are the states where no single site is doubly occupied by intra-pairing atoms, provided that $\left|U_{c,f}\right|>3/2 \, |U_{cf}|$. Let $\hat{G}$ be the projector on this space and $\hat{P}=1-\hat{G}$.

The lowest order correction to $E$ comes at the second order, from
the virtual process consisting in the interchange of 
location of two particles at nearest-neighbour distance. The calculation
simplifies once we note that $\hat{P} H_{1}= H_{1}$ and
that $H_{1}\left|\phi\right\rangle$ is an eigenvector of $H_{0}$ for $\left|\phi\right\rangle$ a ground-state.
The related second order effective Hamiltonian then is found to be
\begin{multline}
H_{eff}=\frac{t^{2}}{4}\underset{\left\langle i,j\right\rangle }{\sum}\left(\frac{1}{\left|U_{c}\right|}\vec{C}_{i}\cdotp\vec{C}_{j}+\frac{1}{\left|U_{f}\right|}\vec{F}_{i}\cdotp\vec{F}_{j}\right)-\\
-\frac{zNt^{2}}{8}\left(\frac{1}{\left|U_{c}\right|}+\frac{1}{\left|U_{f}\right|}\right),
\end{multline}
where $\vec{C}$ and $\vec{F}$ are the associated spin variables
defined by $\vec{S}_{i}=\underset{\sigma\sigma^{\prime}}{\sum}c_{i\sigma}^{\dagger} \vec{\tau}_{\sigma\sigma^{\prime}} c_{i\sigma^{\prime}}$.
The corresponding ground-state energy correction is $\Delta E=-\frac{zNt^{2}}{4}\left(\frac{1}{\left|U_{c}\right|}+\frac{1}{\left|U_{f}\right|}\right)$,
being $z$ the adjacency number for every site. 
In this way the ground-state energy at
the second order perturbation theory in $\frac{t}{U_{c,f}}$ becomes
\begin{equation}
E= -N |U_{cf}| -\frac{zNt^{2}}{4}\left(\frac{1}{\left|U_{c}\right|}+\frac{1}{\left|U_{f}\right|}\right).
\end{equation}
This formula appears in Eq. \eqref{eg1} in the main text.

\section{Strongly coupled $f$ and weakly coupled $c$}
\label{appc}

In this case the system is described by the Hamiltonian
(in the same notation of Appendix \ref{appb}) $H_0+ H_1+H_2$, with:
\begin{equation}
H_{0}=  2 |U_{f}| \underset{i}{\sum}\hat{n}_{iu}\hat{n}_{id} - 2 |U_{cf}| \underset{i,c,f}{\sum}\hat{n}_{ic}\hat{n}_{if}-t\underset{\left\langle i,j\right\rangle ,c}{\sum}c_{ic}^{\dagger}c_{jc}, 
\end{equation}
\begin{equation}
H_{1}=-t\underset{\left\langle i,j\right\rangle, f} \sum  \, c_{i f}^{\dagger}c_{j f} \, ,
\quad \quad H_{2}= -2 U_c \underset{i}{\sum}n_{ir}n_{ig},
\end{equation}
and the perturbative parameters are $\varepsilon_{1}=\frac{t}{|U_{f}|}$ 
and  $\varepsilon_{2}=\frac{|U_{c}|}{t}$.
The ground-state of $H_{0}$ can be derived in this
limit assuming a basis of localized
$f$ degrees of freedom. Using such a basis, we can get an effective Hamiltonian
for the $c$ degrees of freedom corresponding to non-interacting fermions in a
one body potential, in turn depending on the $f$ configuration.

If $\left|U_{f}\right|\gg t$ and $\left|U_{f}\right| \gg |U_{cf}|$, 
the dynamics is dominated by the localization of the $f$ atoms
and therefore the ground-state does not host any doubly occupied site. 
In that case in the ground state of $H_0$, a single $f$ particle 
is in each site, therefore the one-body potential felt by the $c$ particles 
is site independent: $-2\left|U_{cf}\right|\hat{n}_{ic} $. The 
effect of this potential is to induce 
a shift $\delta \mu_c =   -2 \, |U_{cf}|$. 
Up to the first order of perturbation, the ground-state energy then results
of $E_{0c}=2\underset{\vec{k}:\varepsilon_{\vec{k}}<0}{\sum} \, \varepsilon_{\vec{k}} $. Instead the first order in $\epsilon_1$ vanishes because it
is related with forbidden double occupancies of sites 
by particles of the same species. 

At the second order in $\varepsilon_{1}$ and $\varepsilon_{2}$, 
an effective Hamiltonian can be derived:
\begin{multline}
\hat{H}_{eff}=\hat{G}\Bigg[
\varepsilon_{1}^{2}\hat{H}_{1}\frac{1}{\left(E_{0}-\hat{H}_{0}\right)}\hat{P}\hat{H}_{1} +   \\
\varepsilon_{1}\varepsilon_{2}\left(\hat{H}_{1}\frac{1}{\left(E_{0}-\hat{H}_{0}\right)}\hat{P}\hat{H}_{2}+\mathrm{h.c.}\right)+\\
+ \,\varepsilon_{2}^{2}\hat{H}_{2}\frac{1}{\left(E_{0}-\hat{H}_{0}\right)}\hat{P}\hat{H}_{2}\Bigg]\hat{G} \, ,
\end{multline}
wit $\hat{G}$ and $\hat{P}=1-\hat{G}$ as before.
The term $\propto \epsilon_1 \epsilon_2$ vanishes 
for the same reason for which the linear term in $\epsilon_1$ does,
and the remaining effective terms are then proportional to $\varepsilon_{2}$,
$\varepsilon_{1}^{2}$ and $\varepsilon_{2}^{2}$. These 
terms commute with each other, so we can focus on them individually. 
After some algebra
we arrive to the  energy correction up to the second order for the 
ground-state energy:
\begin{equation}
\Delta E=N\left(- \frac{U_{c}}{4}-\frac{zt^{2}}{4}\left|U_{f}\right|-
\frac{U_{c}^{2}}{t}\tilde{E}^{\left(2\right)}\right),
\end{equation}
where $\tilde{E}^{\left(2\right)}$ is a dimensionless positive quantity:
\beq
\tilde{E}^{\left(2\right)}=-\frac{1}{V^{3}}\underset{\begin{array}{c}
\vec{k_{1}},\vec{k_{2}}\in\mathbb{F}_{S}\\
\vec{q_{1}},\vec{q_{2}}\notin\mathbb{F}_{S}
\end{array}}{\sum}\frac{\delta_{\vec{k_{1}}+\vec{k_{2}}\vec{q_{1}}+\vec{q_{2}}}}{\tilde{\varepsilon}_{\vec{k_{1}}}+\tilde{\varepsilon}_{\vec{k_{2}}}-\tilde{\varepsilon}_{\vec{q_{1}}}-\tilde{\varepsilon}_{\vec{q_{2}}}},
\label{delta22}
\eeq 
with $\mathbb{F}_{S}$ labelling the set of points of the Fermi sea and $\tilde{\varepsilon_k}=\varepsilon_k/2t$. 
Eq. \eqref{delta22} is used to arrive to Eq. \eqref{eg2} of the main text, 
where $U_c=0$ and  
it is not needed to calculate $\tilde{E}^{\left(2\right)}$.

\section{Determination of the model parameters}
\label{appA}

In the present Appendix we perform a variational estimate of the 
parameters entering in the Hamiltonian \ref{H}, 
which can be obtained from the expressions 
\begin{equation}
\begin{gathered}
t_{ij\alpha}=-\int d^{3}\vec{r} \,
\Bigg(\frac{\hbar^{2}}{2m_{\alpha}}\nabla\phi_{i\alpha}\left(\vec{r}\right)\cdot\nabla\phi_{j\alpha}\left(\vec{r}\right)
+\\
+ \,  \phi_{i\alpha}\left(\vec{r}\right)V_{\mathrm{ext}}\left(\vec{r}\right)\phi_{j\alpha}\left(\vec{r}\right) \Bigg), \\
{}\\
U_{\alpha\beta}= - \frac{\pi\hbar^{2}a_{\alpha\beta}}{m_{\alpha\beta}}\int d^{3}\vec{r}\left|\phi_{\alpha}\left(\vec{r}\right)\right|^{2}\left|\phi_{\beta}\left(\vec{r}\right)\right|^{2}.
\end{gathered}
\end{equation}
$V_{\mathrm{ext}}\left(\vec{r}\right)  = V_{0}\underset{j=1}{\overset{3}{\sum}}\sin^{2}\left(k_{0}r_{i}\right)$ 
is the external potential creating the lattice 
($k_0 = \frac{2 \pi}{a}$, $a$ being the lattice spacing), 
$a_{\alpha\beta}$ correspond to the scattering lengths between
the $\alpha$ and $\beta$ species, and $m_{\alpha\beta}$
are their reduced masses. Moreover the 
$\phi_{\alpha}\left(\vec{r}\right)$ refer to the Wannier functions 
centered on the lattice sites. A simple estimate of these functions
can be obtained by variational approach. 
In particular we consider the following ansatz:
\begin{equation}
\phi_{\alpha}\left(\vec{r}\right)=C_{\alpha}  e^{-\frac{|\vec{r}|^{2}}{2\sigma_{\alpha }^{2}}},
\label{eq:ansatz}
\end{equation}
where $C_{\alpha}=\left(\sqrt{\pi}\sigma_{\alpha}\right)^{-3/2}$ and the coefficients $\sigma_{\alpha }$ are fixed minimizing
the energy per lattice site.
This value can be found as the expectation value of the 
Hamiltonian \eqref{H} acting on the multi-particles 
fermionic state $\Psi_{\alpha} (\vec{r}_1, \dots, \vec{r}_V)$ 
($V$ being the number of lattice sites, at half filling equal to the 
number of $c$ or $f$ atoms) constructed by the Wannier functions. 
In the mean field approximation it reads:
\begin{multline}
\varepsilon=\int \prod_{i=1}^V \, d^{3}\vec{r}_i \Bigg(\underset{\alpha}{\sum}\frac{\hbar^{2}}{2m_\alpha}\left|\nabla\Psi_{\alpha}\right|^{2}+V_{ext}\left|\Psi_{\alpha}\right|^{2}
+\\
+ \, \underset{\beta>\alpha}{\sum}\frac{2 \pi\hbar^{2}a_{\alpha\beta}}{m_{\alpha\beta}}\left|\Psi_{\alpha}\right|^{2}\left|\Psi_{\beta}\right|^{2} \Bigg).
\end{multline}
Using the (approximate) orthogonality of the Wannier 
functions at different lattice sites one obtains:   
\begin{multline}
\varepsilon=\int d^{3}\vec{r} \, \underset{\vec{r}',\alpha}{\sum} \Bigg(n_\alpha \frac{\hbar^{2}}{2m_{\alpha}}\left|\nabla\phi_{\alpha\vec{r}'}\right|^{2}
+n_\alpha V_{ext}\left|\phi_{\alpha\vec{r}'}\right|^{2}+ \\
+ \, \underset{\beta>\alpha}{\sum}n_\alpha n_\beta\frac{g_{\alpha\beta}}{2}\left|\phi_{\alpha\vec{r}'}\right|^{2}\left|\phi_{\beta\vec{r}'}\right|^{2} \Bigg),
\end{multline}
$n_{\{\alpha, \beta\}}$ being the average number of particles of 
$\{\alpha,\beta\}$ per site and 
$g_{\alpha\beta}=\frac{4\pi\hbar^{2}a_{\alpha\beta}}{m_{\alpha\beta}}$. Moreover 
the Wannier functions, centered on the lattice sites labelled by $\vec{r}^{\, \prime}$, depend on the space vector $\vec{r}$ spanning all the lattice.
Using the ansatz in Eq. \eqref{eq:ansatz} one finds 
\begin{multline}
\varepsilon/N=\underset{\alpha}\sum\Bigg[n_\alpha \frac{\hbar^{2}}{2m_{\alpha}}\frac{3}{2\sigma_{\alpha}^{2}}+n_\alpha \frac{3V_{0}}{2}\left(1-e^{-k_0^2\sigma_\alpha^2}\right) +\\
+ \, \underset{\beta>\alpha}{\sum}n_\alpha n_\beta \frac{g_{\alpha\beta}}{2\pi^{3/2}\left(\sigma_{\alpha}^{2}+\sigma_{\beta}^{2}\right)^{3/2}}\Bigg].
\label{enfin}
\end{multline}
Imposing $\frac{\partial \varepsilon}{\partial\sigma_{\mu}}=0$ 
and expressing the parameters in Eq. \eqref{enfin} as adimensional quantities
$\tilde{\sigma}_{\mu}=k_{0}\sigma_{\mu}$, 
$\tilde{V}_{\alpha}=\frac{V_{0}}{E_{R}^{\alpha}}$
and $\tilde{a}_{\alpha\beta}=k_{0}a_{\alpha\beta}$, with $E_{R}^{\alpha}=\frac{\hbar^{2}k_{0}^{2}}{2m_{\alpha}}$, the result is a set of coupled equations:
\begin{equation}
\frac{1}{\tilde{\sigma}_{\mu}^{3}}-\tilde{V}_{\mu}\tilde{\sigma}_{\mu}e^{-\tilde{\sigma}_{\mu}^{2}}+4\underset{\beta\neq\mu}{\sum}n_\beta \left(1+\frac{m_{\mu}}{m_{\beta}}\right)\frac{\tilde{a}_{\mu\beta}\tilde{\sigma}_{\mu}}{\sqrt{\pi}\left(\tilde{\sigma}_{\mu}^{2}+\tilde{\sigma}_{\beta}^{2}\right)^{5/2}} \,=0.
\end{equation}
Solving this set in $\left\{\sigma_\alpha\right\}$, the Hubbard coefficients 
are finally obtained:
\begin{equation}
\begin{array}{c}
t_{\alpha}=-\left[\frac{\hbar^{2}}{2m_{\alpha}}\frac{1}{4\sigma_{\alpha}^{2}}\left(6-\left(\frac{a}{\sigma_{\alpha}}\right)^{2}\right)+\frac{V_{0}}{2}\left(3-e^{-k_{0}^{2}\sigma_{\alpha}^{2}}\right)\right]e^{-\frac{a^{2}}{4\sigma_{\alpha}^{2}}},\\
\\
{}\\
U_{\alpha\beta}=- \frac{\hbar^2a_{\alpha\beta}}{\sqrt{\pi}m_{\alpha\beta}}\frac{1}{\left(\sigma_{\alpha}^{2}+\sigma_{\beta}^{2}\right)^{3/2}}.
\end{array}
\end{equation}

For the case of the $\mathrm{Yb}$ mixture 
the interactions are the same for the species $r,g$ and $u,d$, 
resulting in two equations (for $\tilde{\sigma_{c}}$ and $\tilde{\sigma_{f}}$):
\begin{equation}
\left\{ \begin{array}{c}
\frac{1}{\tilde{\sigma}_{c}^{3}}-\tilde{V}_{c}\tilde{\sigma}_{c}e^{-\tilde{\sigma}_{c}^{2}}+\frac{n_c\tilde{a}_{cc}}{\sqrt{2\pi}\tilde{\sigma}_{c}^{4}}+\left(1+\frac{m_{c}}{m_{f}}\right)\frac{4n_f\tilde{a}_{cf}\tilde{\sigma}_{c}}{\sqrt{\pi}\left(\tilde{\sigma}_{c}^{2}+\tilde{\sigma}_{f}^{2}\right)^{5/2}}=0\\
\\
\frac{1}{\tilde{\sigma}_{f}^{3}}-\tilde{V}_{f}\tilde{\sigma}_{f}e^{-\tilde{\sigma}_{f}^{2}}+\frac{n_f\tilde{a}_{ff}}{\sqrt{2\pi}\tilde{\sigma}_{f}^{4}}+\left(1+\frac{m_{f}}{m_{c}}\right)\frac{4n_c\tilde{a}_{cf}\tilde{\sigma}_{f}}{\sqrt{\pi}\left(\tilde{\sigma}_{c}^{2}+\tilde{\sigma}_{f}^{2}\right)^{5/2}}=0.
\end{array}\right.
\label{eq:self_consistent}
\end{equation}
The solutions are presented in Fig. \ref{param} of the main text for the symmetric case $n_c=n_f \equiv n =1$.

\end{document}